\documentclass[preprint,showpacs,preprintnumbers,amsmath,amssymb]{revtex4}

\usepackage{graphicx}
\usepackage{dcolumn}
\usepackage{bm}

\begin{document}

\preprint{}

\title{On de Broglie's soliton wave function of many particles with finite masses, energies and momenta}

\author{Agung Budiyono}
\affiliation{Institute for the Physical and Chemical Research, RIKEN, 2-1 Hirosawa, Wako-shi, Saitama 351-0198, Japan}

\date{\today}

\begin{abstract}

We consider a mass-less manifestly covariant {\it linear} Schr\"odinger equation. First, we show that it possesses a class of non-dispersive soliton solution with finite-size spatio-temporal support inside which the quantum amplitude satisfies the Klein-Gordon equation with finite {\it emergent} mass. We then proceed to interpret the soliton wave function as describing a particle with finite mass, energy and momentum. Inside the spatio-temporal support, the wave function shows spatio-temporal internal vibration with angular frequency and wave number that are determined by the energy-momentum of the particle as firstly conjectured by de Broglie. Imposing resonance of the internal vibration inside the spatio-temporal support leads to Planck-Einstein quantization of energy-momentum. The first resonance mode is shown to recover the classical energy-momentum relation developed in special relativity. We further show that the linearity of the Schr\"odinger equation allows one to construct many solitons solution through superposition, each describing a particle with various masses, energies and momenta. 

\end{abstract}

\maketitle  

\section{ Introduction: relativistic Madelung fluid} 

Surprisingly, despite of the remarkable pragmatical successes of quantum theory, there is still an unsettled debate concerning the physical status of its main ingredient, that is the wave function. Roughly speaking, there are two main attitudes toward this foundational issue \cite{Penrose book,Isham book,Bell unspeakable,Bohm-Hiley book}. 

The first cult considers that the wave function has no physical reality at all. They assume the wave function as the representation of {\it our knowledge} about the physical reality rather than to refer to the reality  itself. This interpretation then regards quantum theory as a theory concerning our knowledge about the reality obtained through experiment rather than a theory about physical reality. It is developed by Bohr and Heisenberg and commonly known as Copenhagen interpretation. This line of thought eventually led to the {\it probabilistic view} of wave function through Born's rule \cite{Born paper}. 

The second attitude is to consider the wave function as a real {\it physical field} referring directly to the physical object being described, like say electromagnetic field. There are many interpretation of quantum theory which attribute such a physical status to the wave function. The mostly mentioned interpretations which support this view includes the axiomatic-most-``used'' standard interpretation of Dirac-von Neumann \cite{Dirac book,von Neumann book}, the Bohmian mechanics \cite{Bohm-Hiley book,Bohm paper}, many worlds interpretation \cite{Everett many worlds,deWitt and Graham book}, the theory of spontaneous localization \cite{GRW}, etc. 

The less mentioned one is de Broglie's theory of double solutions \cite{de Broglie book,de Broglie late book}. In his attempt to solve the dual nature of matter as particle and wave, he was searching for a {\it nonlinear wave equation} which assumes a non-dispersive soliton solution at the amplitude which is sufficiently high, while possesses a linear solution satisfying a linear superposition at the amplitude which is weak. He then proposed that the soliton part should be regarded as a particle which is guided by the linear part of the solution. This idea eventually led him to derive his famous guiding principle: 
\begin{equation}
E=\hbar\omega,\hspace{2mm}{\bf p}=\hbar{\bf k},
\label{de Broglie guidance relation}
\end{equation}
which relates the energy-momentum  $\{E,{\bf p}\}$ of the particle with the angular frequency-wave number $\{\omega,{\bf k}\}$ of the linear wave. 

Equations (\ref{de Broglie guidance relation}) can be argued as the most important principle of quantum mechanics \cite{Pauli book}. Partly inspired by those relations, Schr\"odinger developed his celebrated equation. Yet in contrast to de Broglie's original idea, Schr\"odinger equation is linear with respect to the wave function. In this theory, a free particle is then usually represented by a plane wave satisfying the relations  of Eqs. (\ref{de Broglie guidance relation}). Despite physically vague, the plane wave representation surprisingly works for all pragmatical purposes \cite{Bell unspeakable}. Yet one can argue that the successes of this representation relies heavily on  the above de Broglie's relation \cite{de Broglie late book}. It is apparently the pragmatical successes of the linear Schr\"odinger equation if combined with the Born's rule that eventually discourage people from further continuing de Broglie's program \cite{Curfaro-Petroni paper,Vigier paper 1,Vigier paper 2,Mackinnon paper 1,Barut paper}.  In this paper, by considering a mass-less relativistic {\it linear} Schr\"odinger equation, we shall show that it has a new class of soliton solutions with properties exactly envisioned long time ago by de Broglie. 

Let us consider a closed system whose state is uniquely determined by a complex-valued wave function in spacetime, $\psi(q)$, where $q=(q^0,q^1,q^2,q^3)=(ct,x,y,z)$. Further, let us  assume that the wave function satisfies the following manifestly covariant mass-less Schr\"odinger equation
\begin{equation}
i\hbar\frac{\partial}{\partial\lambda}\psi(q;\lambda)=\frac{\hbar^2}{2}\Box\psi(q;\lambda).
\label{covariant Schroedinger equation}
\end{equation}
Here, $\lambda$ is some affine parameter, $\Box=-\eta^{ab}\partial_a\partial_b$ and $\eta^{ab}=\mbox{diag}(-1,1,1,1)$ are D'Alembertian operator and flat Minkowskian metric, respectively. Eq. (\ref{covariant Schroedinger equation}) has been proposed to interpret the Klein-Gordon equation through a particle model \cite{Nambu,Kyprianidis}. 

Next, putting the wave function into polar form, $\psi=I\exp(iS/\hbar)$, where $I$ and $S$ are real-valued functions, and separating into the real and imaginary parts, one obtains \cite{Bohm-Hiley book}:
\begin{eqnarray}
\frac{dv^a}{d\lambda}=-\partial^aU,\hspace{2mm}
\frac{\partial\rho}{\partial\lambda}+\partial_a\big(\rho\hspace{1mm}v^a\big)=0.
\label{covariant Madelung fluid dynamics}
\end{eqnarray}
Here, $\rho(q;\lambda)=|\psi|^2=I^2$ is the quantum probability density, $v^a(q;\lambda)$ is a velocity field generated by the quantum phase $S(q;\lambda)$ as 
\begin{equation}
v^a(q;\lambda)=\partial^aS(q;\lambda),
\label{four velocity vector field}
\end{equation}
and $U(q;\lambda)$ is generated by the quantum amplitude $I(q;\lambda)$ as 
\begin{equation}
U(q)=\frac{\hbar^2}{2}\frac{\Box I}{I}.
\label{covariant quantum potential}
\end{equation}
We have thus adopted the Madelung fluid picture for the Schr\"odinger equation \cite{Madelung paper}. Due to its formal similarity with the Euler equation in hydrodynamics, the term on the right hand side of the left equation in Eqs. (\ref{covariant Madelung fluid dynamics}), $F^a=-\partial^aU$, is called as quantum force field.  Thus, correspondingly, $U(q)$ is called as quantum potential.

Let us remark that written in the form of Madelung fluid, it becomes clear that the original Schr\"odinger equation possesses a hidden self-referential property. Namely, the quantum potential $U(q)$ is generated by the quantum probability density $\rho(q)$ through Eq. (\ref{covariant quantum potential}). This in turn will dictate the way $\rho(q)$ must evolve with time through Eqs. (\ref{covariant Madelung fluid dynamics}) and so on and so forth. It is thus  reasonable to expect some interesting {\it self-organized} physically relevant phenomena. In particular, in this paper we shall be interested to study the {\it fixed points} of such dynamics. 

\section{Self-trapped quantum probability density} 

Let us proceed to specify a class of wave functions whose quantum probability density is further related to its own quantum potential as \cite{AgungPRA1}:
\begin{equation}
\rho(q)=\frac{1}{Z(T)}\exp\Big(-\frac{U(q)}{T}\Big),
\label{canonical QPD}
\end{equation}
where $T$ is a real-valued parameter below chosen to be non-negative and $Z(T)$ is a normalization factor. We shall show that the above class of quantum probability densities possesses non-trivial and physically interesting properties. To do this, notice that combined with the definition of quantum potential given in Eq. (\ref{covariant quantum potential}), Eq. (\ref{canonical QPD}) comprises a differential equation for $U(q)$ or $\rho(q)$ subjected to the condition that $\rho(q)$ must be normalized. In term of $U(q)$, one has to solve the following nonlinear differential equation \cite{AgungPRA1}:
\begin{equation}
-\Box U=\frac{1}{2T}\partial^aU\partial_aU+\frac{4T}{\hbar^2}U. 
\label{covariant NPDE for U}
\end{equation}
One observes that the above differential equation is invariant under Lorentz transformation. Hence, given a solution $U(q)$, then any function $U(q')$, where $q'^a=\Lambda^a_{\hspace{1mm}b}q^b$ and $\Lambda^a_{\hspace{1mm}b}$ is Lorentz transformation, is also a solution of Eq. (\ref{covariant NPDE for U}). 

Let us develop a class of solutions in which the quantum probability density is being trapped by the quantum potential it itself generates \cite{AgungPRA1}. To do this, let us assume that there is an inertial frame so that the quantum probability density is separable into its spatial and temporal parts as follows:
\begin{equation}
\rho(q)=\rho_{\bf x}({\bf x})\rho_t(t),\hspace{2mm}{\bf x}=\{x,y,z\}.
\label{separability}
\end{equation}
In this case, the quantum potential can then be decomposed into 
\begin{equation}
U(q)=U_{\bf x}({\bf x})+U_t(t), 
\label{space-time decomposability}
\end{equation}
where
\begin{equation}
U_{\bf x}({\bf x})=-\frac{\hbar^2}{2}\frac{\partial_{\bf x}^2I_{\bf x}}{I_{\bf x}},\hspace{2mm}U_t(t)=\frac{\hbar^2}{2c^2}\frac{\partial_t^2I_t}{I_t}.
\label{decomposable quantum potential}
\end{equation}
Here, $I_i\equiv\sqrt{\rho_i}$ with $i={\bf x},t$; and $\partial_{\bf x}^2\equiv\partial_{\bf x}\cdot\partial_{\bf x}$ where $\partial_{\bf x}=\{\partial_x,\partial_y,\partial_z\}$. 

The condition of Eq. (\ref{separability}) is not Lorentz invariant so is the resulting class of solutions we are going to develop. Yet, its nontrivial property will be shown to be Lorentz invariant. Inserting  Eq. (\ref{space-time decomposability}), Eq. (\ref{covariant NPDE for U}) can thus be re-collected as $\partial_{\bf x}^2U_{\bf x}-(1/2T)\partial_{\bf x}U_{\bf x}\cdot\partial_{\bf x}U_{\bf x}-(4T/\hbar^2)U_{\bf x}
=(1/c^2)\partial_t^2U_t-(1/2T)(\partial_tU_t)^2+(4T/\hbar^2)U_t=D,$
where $D$ is constant. Below for simplicity we shall take the case when $D=0$. One thus has to solve the following decoupled pair of nonlinear differential equations:
\begin{eqnarray}
\partial_{\bf x}^2U_{\bf x}-\frac{1}{2T}\partial_{\bf x}U_{\bf x}\cdot\partial_{\bf x}U_{\bf x}-\frac{4T}{\hbar^2}U_{\bf x}=0,\nonumber\\
\frac{1}{c^2}\partial_t^2U_t-\frac{1}{2T}(\partial_tU_t)^2+\frac{4T}{\hbar^2}U_t=0.\hspace{2mm}
\label{decoupled NPDE}
\end{eqnarray}

\subsection{Spatial self-trapping}

Let us first discuss the spatial part by solving the upper differential equation in Eqs. (\ref{decoupled NPDE}). To do this, let us search for a class of solutions in which the spatial quantum probability density is further separable as 
\begin{equation}
\rho_{\bf x}({\bf x})=\rho_x(x)\rho_y(y)\rho_z(z), 
\label{spatial separability}
\end{equation}
so that the spatial part of the quantum potential is further decomposable into 
\begin{equation}
U_{\bf x}({\bf x})=U_x(x)+U_y(y)+U_z(z), \hspace{2mm}U_i(i)=-\frac{\hbar^2}{2}\frac{\partial_i^2I_i}{I_i},
\label{spatial decomposable quantum potential}
\end{equation}
where $i=x,y,z$. Putting this anzatz into the upper differential equation in Eqs. (\ref{decoupled NPDE}), one  can choose a class of solutions in which each $U_i(i)$ satisfies the following decoupled nonlinear differential equations:
\begin{equation}
\partial_i^2U_i=\frac{1}{2T}(\partial_iU_i)^2+\frac{4T}{\hbar^2}U_i,\hspace{2mm}i=x,y,z. 
\label{NPDE for spatial U}
\end{equation}
To avoid complicated notation, below we shall consider only the $x-$degree of freedom. The other two spatial degrees of freedom $\{y,z\}$ follows similarly. 

\begin{figure}[htbp]
\begin{center}
\includegraphics*[width=6cm]{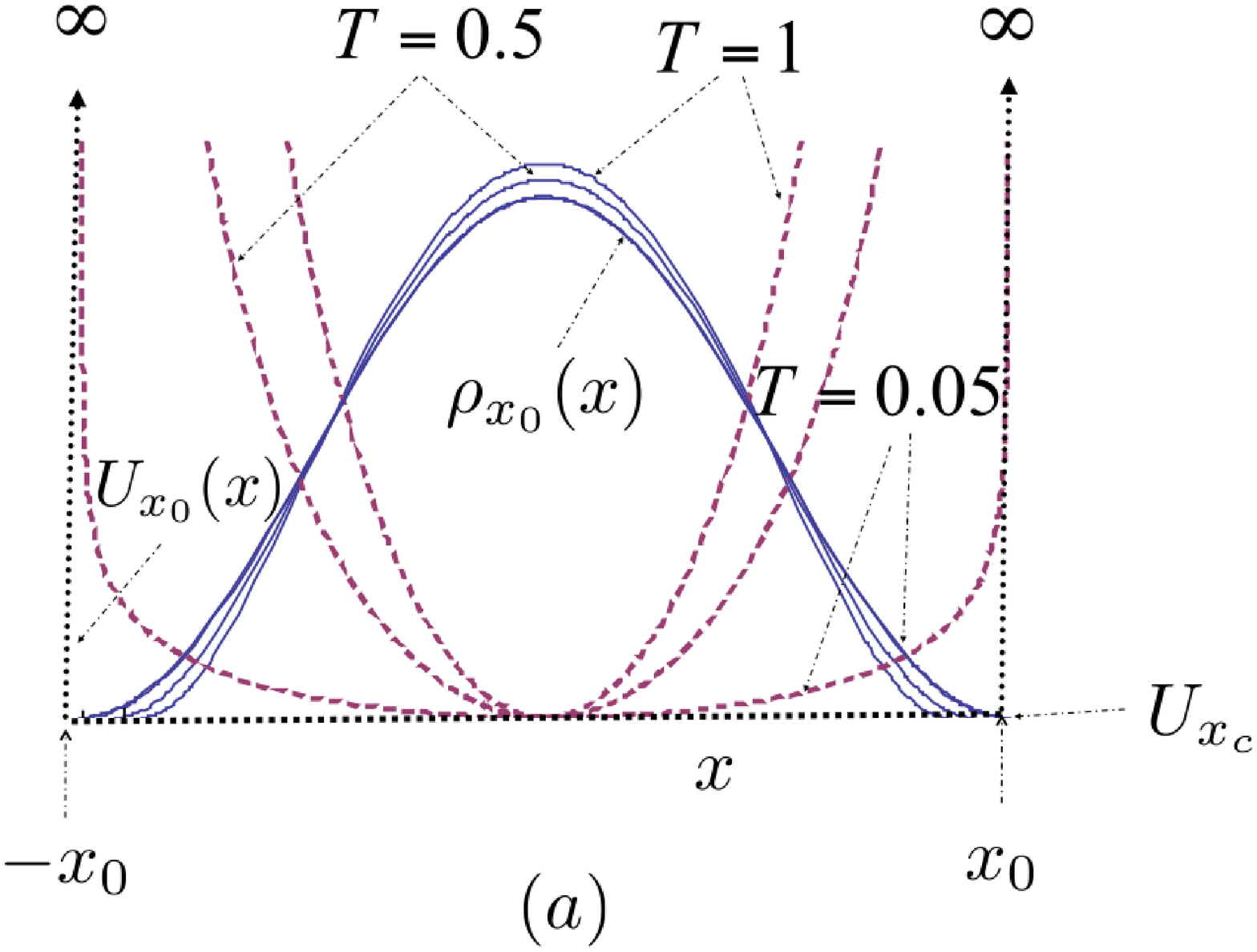}
\includegraphics*[width=7.5cm]{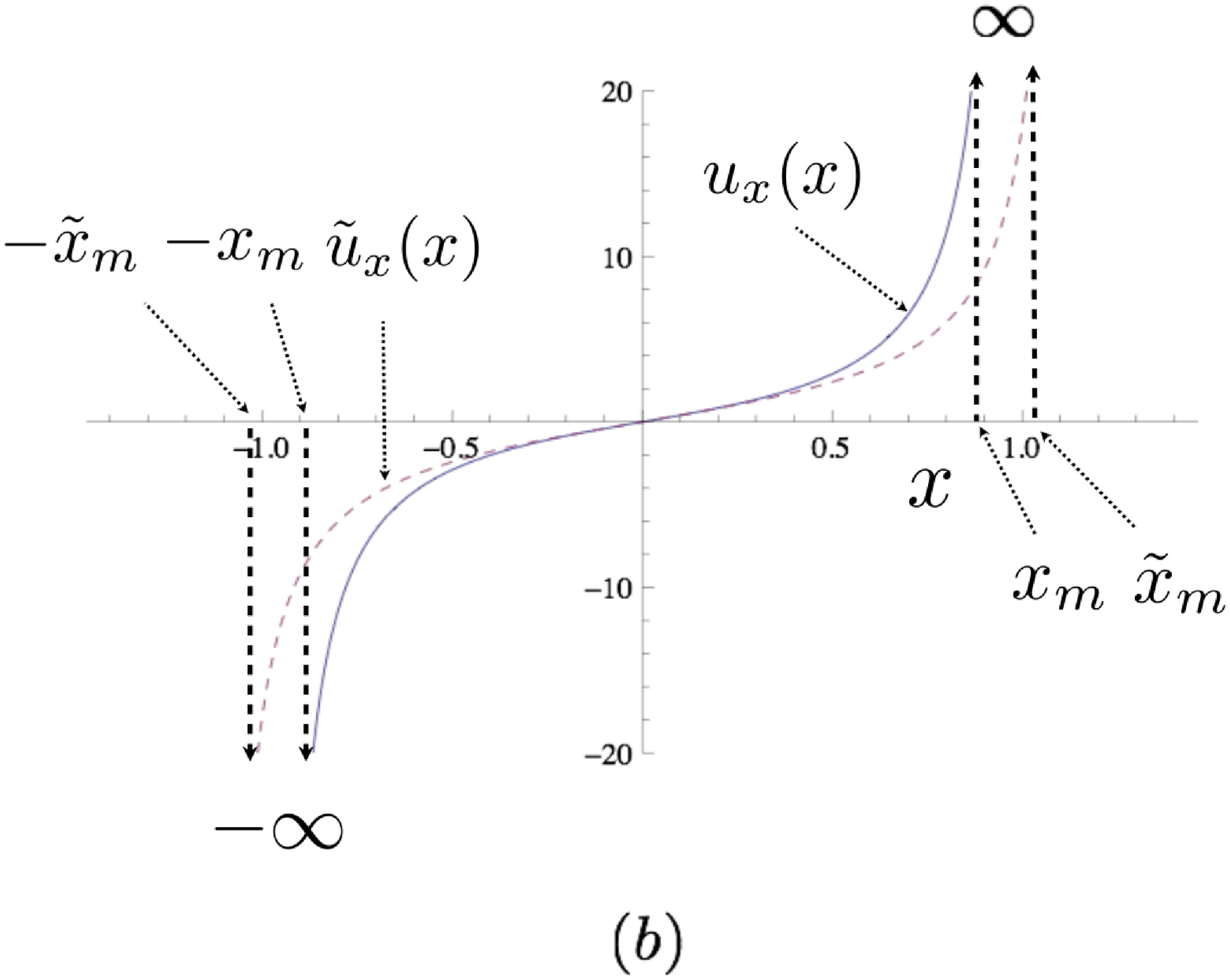}
\end{center}
\caption{(a) The $x-$part of spatial quantum probability density $\rho_x(x)$ (solid line) and its corresponding partial quantum potential (dashed line) $U_x(x)$ for several small values of $T$ obtained by solving equation (\ref{NPDE for spatial U}). We also plot the analytical solution for $\rho_{x_0}(x)$ at $T=0$, assuming that $U_{x_0}(x)$ takes the form of a box with infinite wall. (b) $u_x(x)$ and $\tilde{u}_x(x)$. See text for detail.}
\label{1D QPD-QP}
\end{figure}

Figure \ref{1D QPD-QP}a shows the numerical solutions of Eq. (\ref{NPDE for spatial U})  with the boundary conditions: $\partial_xU_x(0)=0$ and $U_x(0)=1$, for several small values of $T$. The reason for choosing small $T$ will be clear later. One can see that the spatial part of quantum probability density $\rho_x(x)$ is being trapped by its own self-generated quantum potential $U_x(x)$ \cite{AgungPRA1}. Moreover, there is a finite distance $x=\pm x_m$ at which the partial quantum potential is blowing-up $U_x(\pm x_m)=\infty$, so that the corresponding partial quantum probability density is vanishing, $\rho_x(\pm x_m)=0$. This is a familiar phenomena in nonlinear differential equation \cite{blowing-up NDE}, which for the case at hand, can be proven as follows. 

Let us define a new variable $u_x(x)=\partial_xU_x$. The nonlinear differential equation of Eq. (\ref{NPDE for spatial U}) then transforms into 
\begin{equation}
\partial_xu_x=\frac{1}{2T}u_x^2+\frac{4T}{\hbar^2}U_x.
\label{blowing-up NPDE for u}
\end{equation}
The boundary condition translates into $u_x(0)=\partial_xU_x(0)=0$. Further, let us now consider the following nonlinear differential equation
\begin{equation}
\partial_x\tilde{u}_x=\frac{1}{2T}\tilde{u}_x^2+\frac{4T}{\hbar^2}X, 
\label{blowing-up for inferior solution}
\end{equation}
where $X\equiv U_x(0)$; with $\tilde{u}_x(0)=0$. Since $U_x(x)\ge U_x(0)=X$, then it is obvious that $|u_x(x)|\ge|\tilde{u}_x(x)|$. 

\begin{figure}[tbp]
\begin{center}
\includegraphics*[width=6cm]{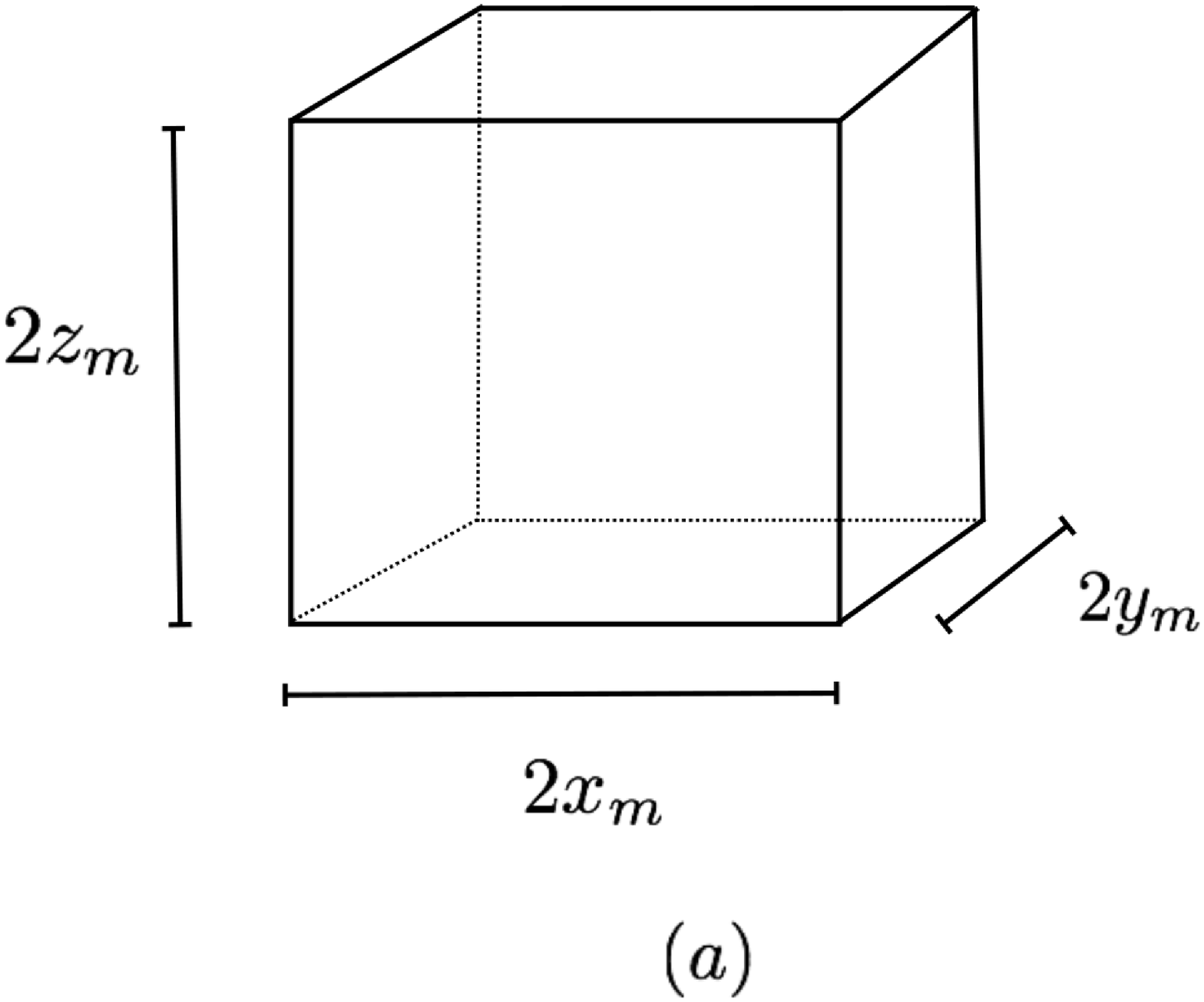}
\includegraphics*[width=6cm]{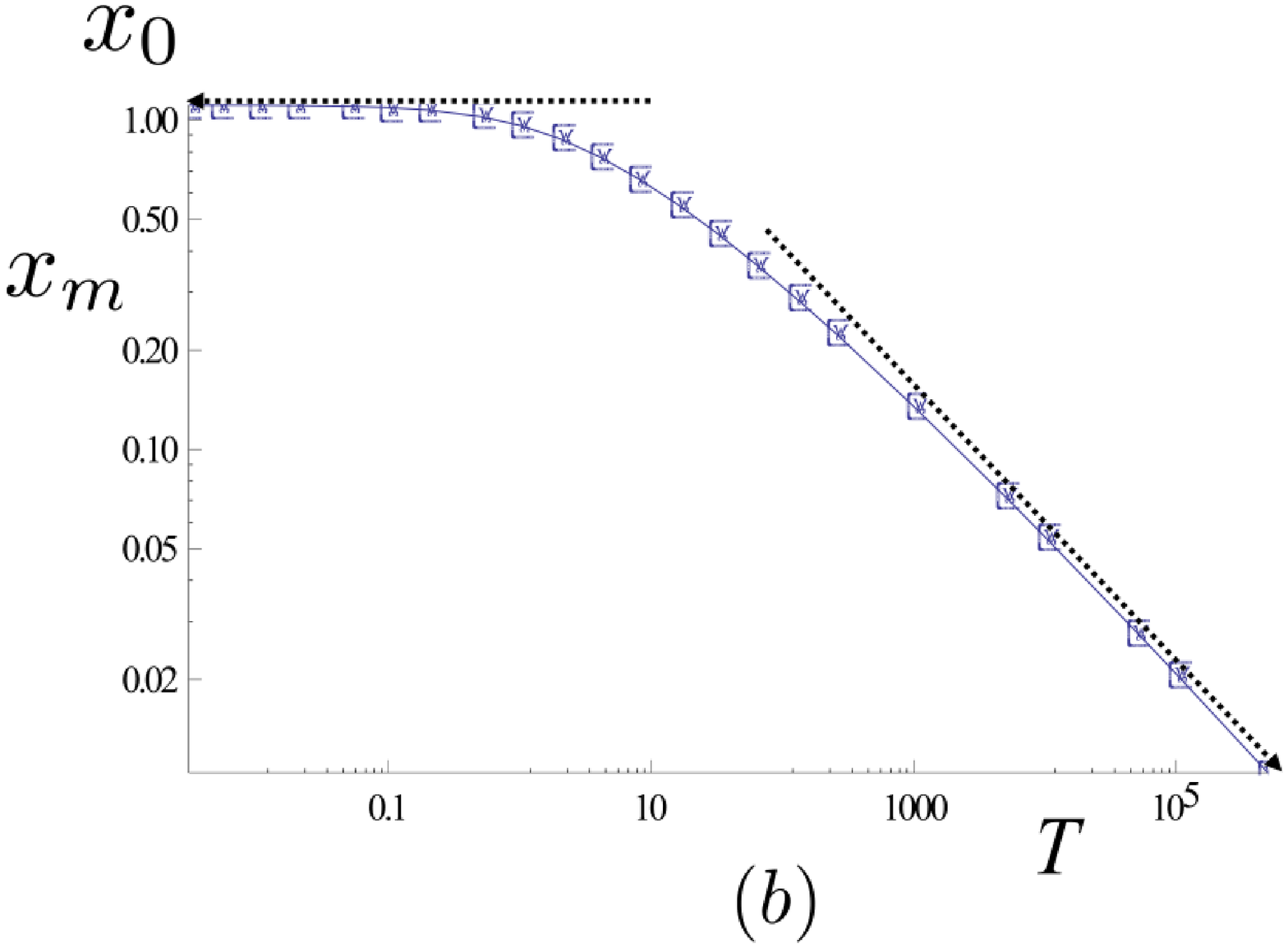}
\end{center}
\caption{(a) The square box spatial support and (b) the half-length of the $x-$side of the box $x_m$ plotted against $T$.}
\label{rmax vs T}
\end{figure}

One can then solve the latter nonlinear differential equation of Eq. (\ref{blowing-up for inferior solution}) analytically to have:
\begin{equation}
\tilde{u}_x(x)=a\tan(bx), \hspace{2mm}a=\frac{2T}{\hbar}\sqrt{2X}, \hspace{2mm}b=\frac{1}{\hbar}\sqrt{2X}.  
\label{inferior solution}
\end{equation}
It is then clear that at $x=\pm\tilde{x}_m=\pm\pi/(2b)$, $\tilde{u}_x$ is blowing-up, namely $\tilde{u}_x(\pm\tilde{x}_m)=\pm\infty$. Recalling the fact that $|u_x(x)|\ge|\tilde{u}_x(x)|$, then $u_x(x)$ is also blowing-up at points $x=\pm x_m$, $u_x(\pm x_m)=\pm\infty$, where $x_m\le \tilde{x}_m$. See Fig. \ref{1D QPD-QP}b. Hence, one can conclude that $U_x(x)$ is also blowing-up at $x=\pm x_m$, $U_x(\pm x_m)=\infty$. It is then safe to say that the $x-$part of spatial self-trapped quantum probability density $\rho_x(x)$ possesses only a finite range of spatial support: $\mathcal{M}_x=[-x_m,x_m]$.  Finally, the spatial part self-trapped quantum probability density $\rho_{\bf x}({\bf x})$ possesses only finite-size spatial support: $\mathcal{M}_{\bf x}=[-x_m,x_m]\otimes[-y_m,y_m]\otimes[-z_m,z_m]$, which takes the form of a three dimensional square box with sides length $2i_{m}$, $i=x,y,z$. See Fig. \ref{rmax vs T}a. 

Let us see what happens if one varies the parameter $T$. Figure \ref{rmax vs T}b shows the values of $x_m$ as a function of $T$ obtained by numerically solving the differential equation of Eq. (\ref{NPDE for spatial U}) with fixed boundary conditions: $\partial_xU_x(0)=0$, $U_x(0)=1$. One can see that as we increase $T$, $x_m$ decreases, and eventually vanishes for infinite value of $T$. This shows that $\rho_x(x)$ is converging toward a delta function for infinite $T$. A very interesting fact is seen for the opposite limit of vanishing $T$. One observes that $\lim_{T\rightarrow 0}x_m(T)=x_0$, where $x_0$ is finite. This fact suggests to us that at $T=0$, the spatial quantum probability density and thus its corresponding quantum potential are converging toward certain functions:
\begin{equation}
\lim_{T\rightarrow 0}\rho_{x}(x;T)=\rho_{x_0}(x),\hspace{2mm}\lim_{T\rightarrow 0}U_x(x;T)=U_{x_0}(x).
\label{frozen spatial QPD and QP}
\end{equation}

Let us discuss this latter asymptotic situation in more detail. First, one can see in Fig. \ref{1D QPD-QP}a that as $T$ decreases, the quantum potential inside the spatial support is getting flatterer before becoming infinite at the boundary points, $x=\pm x_m(T)$. One might then guess that  at $T=0$, the quantum potential is perfectly flat inside the one dimensional box of spatial support and is infinite at its boundary points: $x=\pm x_0$. Guided by this guess, let us calculate the profile of the spatial quantum probability density for vanishing value of $T$. To do this, let us denote the assumed positive definite constant value of the quantum potential inside the support as $U_{x_c}$. Recalling the definition of spatial quantum potential $U_x(x)$ given in Eq. (\ref{spatial decomposable quantum potential}) one has  
\begin{equation}
\partial_x^2 I_{x_0}(x)=-\frac{2U_{x_c}}{\hbar^2}I_{x_0}(x),
\label{Schroedinger equation for a frozen box}
\end{equation}
where $I_{x_0}\equiv\sqrt{\rho_{x_0}}$. The above differential equation must be subjected to the spatial boundary condition: $\rho_{x_0}(\pm x_0)=I_{x_0}^2(\pm x_0)=0$. Solving Eq. (\ref{Schroedinger equation for a frozen box}) one has 
\begin{equation}
I_{x_0}(x)=A_{x_0}\cos(k_{x_0}x),
\label{quantum amplitude for a frozen box}
\end{equation}
where $A_{x_0}$ is normalization constant and the wave number $k_{x_0}$ is related to the quantum potential as:
\begin{equation}
k_{x_0}=\sqrt{2U_{x_c}/\hbar^2}. 
\label{wave number vs quantum potential}
\end{equation}
The boundary condition implies
\begin{equation}
k_{x_0}x_0=\pi/2.
\label{wave number vs support} 
\end{equation}
Figure \ref{1D QPD-QP}a shows that as $T$ decreases toward zero, $\rho_{x}(x;T)$ obtained by numerically solving Eq. (\ref{NPDE for spatial U}) is indeed converging toward $\rho_{x_0}(x)$ obtained in Eq. (\ref{quantum amplitude for a frozen box}). This observation thus confirms our guess that {\it at $T=0$, the $x-$part spatial quantum potential is flat inside the spatial support $\mathcal{M}_x$ and is infinite at its boundary points}. 

Hence, at $T=0$, in total the spatial part of the quantum probability density can be written as
\begin{equation}
\rho_{{\bf x}_0}({\bf x})=\prod_{i=x,y,z}\rho_{i_0}(i), 
\label{spatial 3D self-trapped QPD}
\end{equation}
where $\rho_{i_0}=A_{i_0}\cos(k_{i_0}i)$, $i=x,y,z$. Moreover, the  support is given by three dimensional volume $\mathcal{M}_{{\bf x}_0}=[-x_0,x_0]\otimes[-y_0,y_0]\otimes[-z_0,z_0]$ at surface of which the spatial quantum potential is blowing-up so that the quantum probability density is vanishing. 

\subsection{Temporal self-trapping}

\begin{figure}[tbp]
\begin{center}
\includegraphics*[width=6cm]{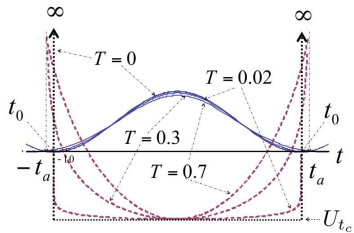}
\end{center}
\caption{The profile of temporal self-trapped quantum probability density $\rho_t(t)$ (solid line) and its corresponding temporal quantum potential $U_t(t)$ (dashed line) for several small values of $T$. See text for detail.}
\label{tempo-creation-annihilation}
\end{figure}

Next, let us discuss the temporal part of quantum probability density $\rho_t(t)$ by solving the lower differential equation in Eqs. (\ref{decoupled NPDE}). In particular, we are interested to investigate the behavior of $\rho_t(t)$ at the limit $T\rightarrow 0$, if it exists. Figure \ref{tempo-creation-annihilation} shows the solution with the boundary: $\partial_tU_t(0)=0$ and $U_t(0)\equiv U_{t_c}=-1$. $\rho_t(t)$ and the corresponding $U_t(t)$ are plotted for several small values of $T$. One can again see similar phenomena with the spatial part that $\rho_t(t)$ is being self-trapped by the corresponding $U_t(t)$. One also sees that the support of $\rho_t(t)$ is finite given by the interval $\mathcal{M}_t=[-t_a,t_a]$ at the boundary points of which the temporal quantum potential is blowing-up: $U_t(\pm t_a)=\infty$. 

\begin{figure}[htbp]
\begin{center}
\includegraphics*[width=7.5cm]{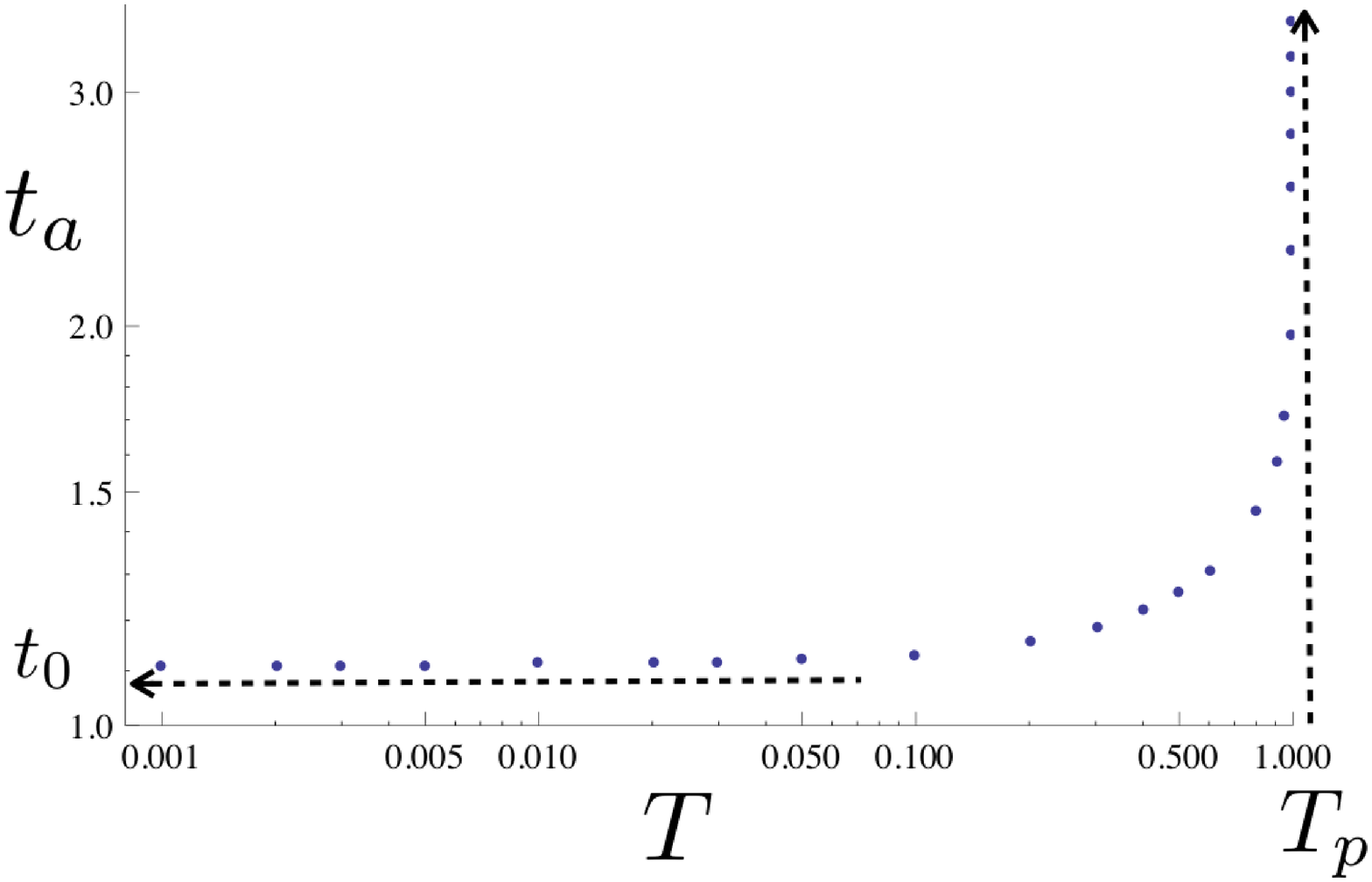}
\end{center}
\caption{The variation of half support $t_a$ of the self-trapped temporal quantum probability density $\rho_t(t)$ against T. See text for detail.}
\label{tmax vs T}
\end{figure}

\begin{figure}[tbp]
\begin{center}
\includegraphics*[width=9cm]{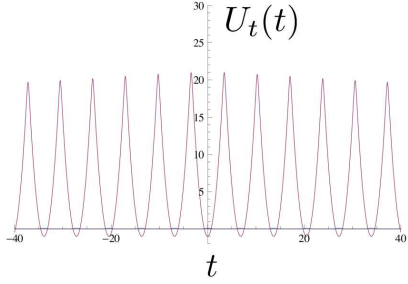}
\end{center}
\caption{Periodic temporal quantum potential for the case $T=1$ with $\partial_tU_t(0)=0$ and $U_t(0)=-1$. See text for detail.}
\label{periodic temporal quantum potential}
\end{figure}

In Fig. \ref{tmax vs T} we plot the variation of $t_a$ against $T$ while keeping $U_t(0)\equiv U_{t_c}$ fixed. First, in contrast to the spatial part, one observes that $t_a$ is a monotonically increasing function of $T$. Further, in contrast to the spatial part in which the length of the support is vanishing for infinite value of $T$, the length of the support of temporal quantum probability density $\rho_t(t)$ is blowing-up at finite value of $T=T_p(U_{t_c})$. See Fig. \ref{tmax vs T}. Given $U_{t_c}$, then for $T\ge T_p(U_{t_c})$, the temporal quantum potential $U_t(t)$ is no more convex everywhere, but is almost periodic as depicted in Fig. \ref{periodic temporal quantum potential}. Hence, $\rho_t(t)$ as defined in Eq. (\ref{canonical QPD}) is no more normalizable. This case is therefore physically irrelevant. 

Yet, again, as in the case of spatial quantum probability density, as one decreases $T$ toward zero,  $t_a$ is converging toward a finite value $t_0$:
\begin{equation}
\lim_{T\rightarrow 0}t_a\equiv t_0. 
\end{equation}
This again shows that at $T=0$, $U_t(t)$ and $\rho_t(t)$ will converge toward some functions:
\begin{equation}
\lim_{T\rightarrow 0}U_t(t;T)=U_{t_0}(t),\hspace{2mm}\lim_{T\rightarrow 0}\rho_t(t;T)\equiv\rho_{t_0}(t). 
\label{converging temporal QPD-QP for vanishing T}
\end{equation}
Below we shall be interested to further study the case of vanishing $T$. 

Proceeding in the same way as for the spatial part, let us calculate $\rho_{t_0}(t)$. To do this, first one observes in Fig. \ref{tempo-creation-annihilation} that as $T$ is approaching zero, $U_t(t)$ is getting flatterer inside the support before becoming infinite at the boundary points, $t=\pm t_a(T)$. Again, let us guess that at $T=0$, the temporal quantum potential is perfectly flat inside the support $\mathcal{M}_{t_0}=(-t_0,t_0)$, given by $U_{t_c}<0$; and is infinite at $t=\pm t_0$. Recalling the definition of $U_t(t)$ given in Eq. (\ref{decomposable quantum potential}), one has
\begin{equation}
\partial_t^2I_{t_0}=\frac{2c^2U_{t_c}}{\hbar^2}I_{t_0}.
\label{Schroedinger equation for time}
\end{equation}
Here $I_{t_0}\equiv\sqrt{\rho_{t_0}}$. The above differential equation must be subjected to the boundary condition: $I_{t_0}(\pm t_0)=0$. Solving Eq. (\ref{Schroedinger equation for time}), one obtains:
\begin{eqnarray}
I_{t_0}(t)=A_{t_0}\cos(\omega_0 t). 
\label{temporal lowest state}
\end{eqnarray}
Here $A_{t_0}$ is a normalization constant and the angular frequency $\omega_0$ is related to the quantum potential as 
\begin{equation}
\omega_0=\sqrt{(-2c^2U_{t_c}/\hbar^2)}. 
\label{angular frequency vs quantum potential}
\end{equation}
The boundary imposes: 
\begin{equation}
\omega_0t_0=\pi/2.
\label{angular frequency vs support} 
\end{equation}
One finally sees in Fig. \ref{tempo-creation-annihilation} that  as one decreases $T$ toward zero, $\rho_t(t;T)$ obtained by solving the lower differential equation in Eqs. (\ref{decoupled NPDE}) is indeed converging toward $\rho_{t_0}(t)$ given in Eq. (\ref{temporal lowest state}). {\it This again justifies our guess that at $T=0$, $U_{t}(t)$ is perfectly flat inside $\mathcal{M}_{t_0}$, and is infinite at $t=\pm t_0$}.   

\section{Spacetime soliton}

Hence, in total, at $T=0$, $\rho_0(q)\equiv\rho_{{\bf x}_0}({\bf x})\rho_{t_0}(t)$ satisfies the differential equation of Eq. (\ref{covariant NPDE for U}). Notice that the spatio-temporal support of $\rho_0(q)$ is  composed by $\mathcal{M}\equiv\mathcal{M}_{{\bf x}_0}\otimes\mathcal{M}_{t_0}$. Inside $\mathcal{M}$, the quantum potential is thus flat given by 
\begin{equation}
U(q)=U_{x_c}+U_{y_c}+U_{z_c}+U_{t_c}=\frac{1}{2}\Big(\hbar^2k_{\bf x_0}^2-\frac{\hbar^2\omega_0^2}{c^2}\Big)\equiv U_0, 
\label{quantum potential-energy-momentum}
\end{equation}
where   $k_{{\bf x}_0}^2={\bf k}_{{\bf x}_0}\cdot {\bf k}_{{\bf x}_0}$ and ${\bf k}_{{\bf x}_0}\equiv \{k_{x_0},k_{y_0},k_{z_0}\}$; and we have employed Eqs. (\ref{wave number vs quantum potential}) and (\ref{angular frequency vs quantum potential}). One therefore observes that at $T=0$, the quantum force is vanishing inside the spatio-temporal support, $\partial^aU=0$. 

\begin{figure}[htbp]
\begin{center}
\includegraphics*[width=7.5cm]{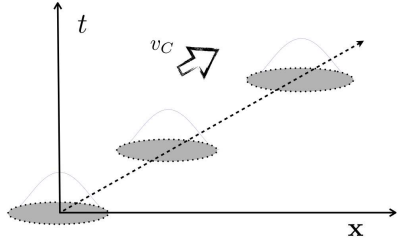}
\end{center}
\caption{Spatio-temporally localized wave packet moving in spacetime.}
\label{spacetime soliton}
\end{figure}

Now, at $\tau=0$, let us choose the following initial wave function, $\{\rho_0(q),v_0^{\hspace{1mm}a}(q)\}$. Here $\rho_0(q)=\rho_{{\bf x}_0}({\bf x})\rho_{t_0}(t)$ and $v_0^{\hspace{1mm}a}(q)$ is a uniform velocity vector field having non-vanishing value only inside the spatio-temporal support $\mathcal{M}$, that is $v_0^{\hspace{1mm}a}(q)=v_C^{\hspace{1mm}a}$, where $v_C^{\hspace{1mm}a}$ is a constant four velocity vector. Since at $\tau=0$ the quantum force is vanishing, then initially one has $dv^a/d\tau=0$. Hence, at infinitesimal lapse of proper time, $\tau=\Delta\tau$, the velocity field is kept uniform and constant. This in turn will shift the initial quantum probability density in spacetime by $\Delta q^{a}=v_C^{\hspace{1mm}a}\Delta\tau$, while keeping its profile unchanged: $\rho(q;\Delta \tau)=\rho_0(q^a-v_C^{\hspace{1mm}a}\Delta \tau)$. Accordingly, the spatio-temporal support will also be shifted by the same amount: $\mathcal{M}_{\Delta\tau}$. The same thing will happen for the next infinitesimal lapse of proper time and so on and so forth. Hence, at finite lapse of proper time $\tau$, one concludes that the pair of fields  
\begin{equation}
\{\rho(q;\tau),v^a(q;\tau)\}=\{\rho_0(q^b-v_C^{\hspace{1mm}b}\tau),v_0^{\hspace{1mm}a}(q^b-v_C^{\hspace{1mm}b}\tau)\},
\end{equation}
comprises the stationary wave function of the relativistic Madelung fluid dynamics. Here, $q$ belongs to the spatio-temporal support at proper time $\tau$, denoted by  $\mathcal{M}_{\tau}$. {\it We have thus a spatio-temporally localized wave packet traveling in spacetime: namely a spacetime soliton}. See Fig. \ref{spacetime soliton}. 

\subsection{Mass} 

Next, before proceeding to write down the explicit form of the spacetime soliton wave function, let us first discuss the physical meaning of quantum potential. First, since the quantum potential is constant inside $\mathcal{M}_{\tau}$ given by $U_0$, one has ${\bar U}=\int dq U\rho=U_0$. Let us proceed to choose a sufficiently large $\omega_0$ by picking sufficiently large $|U_{t_c}|$ so that ${\bar U}=U_0$ given in Eq. (\ref{quantum potential-energy-momentum}) is negative, ${\bar U}<0$. This allows us to define a new quantity $m$ as 
\begin{equation}
{\bar U}=U_0=-\frac{1}{2}m^2c^2, 
\label{emergent mass}
\end{equation}
so that inserting into Eq. (\ref{quantum potential-energy-momentum}) one obtains
\begin{equation}
\frac{\hbar^2\omega_0^2}{c^2}-\hbar^2k_{{\bf x}_0}^2=m^2c^2.
\label{energy-momentum-mass relation}
\end{equation}
Recalling again the definition of quantum  potential of Eq. (\ref{covariant quantum potential}), Eq. (\ref{emergent mass}) can be rewritten as 
\begin{equation}
\Box I(q;\tau)+\frac{m^2c^2}{\hbar^2}I(q;\tau)=0, \hspace{3mm}q\in\mathcal{M}_{\tau},
\label{Klein-Gordon equation}
\end{equation}
where $I\equiv\sqrt{\rho}$. 

A physical interpretation to the above formalism is in order. Eq. (\ref{Klein-Gordon equation}) is but the Klein-Gordon equation with mass term $m$. One however should keep in mind that the differential equation of Eq. (\ref{Klein-Gordon equation}) must be subjected to the boundary condition that $I(q;\tau)$ is vanishing at the surface boundary of the spatio-temporal support, $\mathcal{M}_{\tau}$. One can also see that though $I(q)$ is not Lorentz invariant, $m$ is. Moreover, since $U_0$ is conserved so is $m$. 

\subsection{Energy-momentum}

Let us proceed to discuss another conserved quantity: $v^a={v_C}^a$. To do this, let us use the conserved invariant quantity $m$ developed in the previous subsection to rescale the affine parameter $\lambda$ as follows: 
\begin{equation}
\tilde{\lambda}=m\lambda. 
\label{mass-scaled affine parameter}
\end{equation}
One thus has $\partial/\partial\lambda=m\partial/\partial\tilde{\lambda}$. Using this, Eq. (\ref{covariant Schroedinger equation}) becomes
\begin{equation}
i\hbar\frac{\partial\psi}{\partial{\tilde{\lambda}}}=\frac{\hbar^2}{2m}\Box\psi. 
\label{covariant Schroedinger equation with mass}
\end{equation}
Moreover, the pair of equation in Eqs. (\ref{covariant Madelung fluid dynamics}) translates into
\begin{equation}
m\frac{d{\tilde v}^a}{d{\tilde \lambda}}=-\partial^a{\tilde U},\hspace{2mm}
\frac{\partial\rho}{\partial{\tilde \lambda}}+\partial_a\big(\rho\hspace{1mm}{\tilde v}^a\big)=0,
\label{covariant Madelung fluid dynamics with mass}
\end{equation}
where ${\tilde v^a}\equiv dq^a/d{\tilde \lambda}$, and is now related to the quantum phase $S(q)$ as
\begin{equation}
{\tilde v}^a=\partial^aS/m=v^a/m,
\label{rescaled guiding relation}
\end{equation}
and ${\tilde U}$ is given by 
\begin{equation}
{\tilde U}=U/m=\frac{\hbar^2}{2m}\frac{\Box I}{I}.
\label{rescaled quantum potential}
\end{equation}

Let us then use the proper time $\tau$ for the affine parameter, $\tau=\tilde{\lambda}$, and to avoid complicated notation let us rewrite the rescaled four velocity vector ${\tilde v}^a$ back as $v^a$.  One can then define four momentum vector as 
\begin{equation}
p^a=mv^a. 
\label{four momentum vector}
\end{equation}
which is conserved. The spatial part of the above quantity gives us the usual definition of classical momentum:
\begin{equation}
{\bf p}=m{\bf v}. 
\label{classical momentum}
\end{equation}
From the temporal part one can define a scalar quantity $E$ as
\begin{equation}
E=cp^0=\frac{mc^2}{\sqrt{1-(v/c)^2}}\rightarrow mc^2+\frac{1}{2}mv^2, 
\end{equation}
where $v$ is the absolute value of spatial velocity vector, $v=\|{\bf v}\|$, and the arrow implies that $v/c$ is sufficiently small. Hence, $E$ has the dimension of energy. The second term is the kinetic energy and the first term is usually dubbed as rest-mass energy, namely the energy when the soliton is not moving. 

Next, let us calculate a quantity which is convenient for later discussion defined as follows:
\begin{equation}
\langle P\rangle=\int_{\mathcal{M}} dq\hspace{1mm}\psi^*(q)\Big(\frac{\hbar^2}{2m}\Box\Big)\psi(q).
\label{what is this?}
\end{equation}
Since it is conserved, then it is sufficient to calculate its value at $\tau=0$. Putting the wave function in polar form $\psi=I_0\exp(iS_0/\hbar)$, one gets
\begin{eqnarray}
\langle P\rangle=\int_{\mathcal{M}}dq\hspace{1mm}\Big(I_0\frac{\hbar^2}{2m}\Box I_0+\frac{1}{2m}I_0^2\partial_aS_0\partial^aS_0\hspace{5mm}\nonumber\\
-\frac{i\hbar}{m} I_0\partial_aI_0\partial^aS_0+i\frac{\hbar^2}{2m}I_0^2\Box S_0\Big). 
\label{calculation of what it is}
\end{eqnarray}
The first term on the right hand side is nothing but the average quantum potential divided by mass $m$:
\begin{equation}
\int_{\mathcal{M}}dq\hspace{1mm}I_0\frac{\hbar^2}{2m}\Box I_0=\frac{1}{m}\int_{\mathcal{M}}dq\hspace{1mm}\rho_0(q)U_0(q)=\frac{{\bar U}_0}{m}. 
\end{equation}
Recalling the relation between the four velocity vector field given in Eq. (\ref{rescaled guiding relation}), the second term on the right hand side is given by
\begin{equation}
{\bar K}_0\equiv \int_{\mathcal{M}}dq\hspace{1mm}\rho_0\big(p_a p^a/(2m)\big)=\frac{p_ap^a}{2m}, 
\label{spacetime kinetic energy}
\end{equation} 
where in the last equality we have used the fact that the velocity vector is uniform inside the spatio-temporal support. Further, for the case of soliton where $v^a=\partial^aS_0/m$ is uniform, the last term is vanishing due to the vanishing divergence $\partial_av^a=-\Box S_0/m=0$. Finally, the third term is also vanishing due to the fact that $I_0(q)=I_{t_0}(t)I_{x_0}(x)I_{y_0}(y)I_{z_0}(z)$ is separable and each $I_{i_0}(i)$ possesses a symmetry $I_{i_0}(i)=I_{i_0}(-i)$ so that  $\partial_iI_{i_0}$ satisfies $\partial_iI_{i_0}(-i)=-\partial_iI_{i_0}(i)$, $i=t,x,y,z$. Hence, in total one obtains
\begin{eqnarray}
\langle P\rangle=\frac{{\bar U}_0}{m}+{\bar K_0}=\frac{1}{2m}\Big(\hbar^2k_{{\bf x}_0}^2-\frac{\hbar^2\omega_0^2}{c^2}+p_a p^a\Big)\nonumber\\
=\frac{\big(-m^2c^2+p^a p_a\big)}{2m}=-mc^2.\hspace{10mm}
\label{it is what is this?}
\end{eqnarray} 
Hence it is given by the negative of the rest-mass energy. 

\subsection{de Broglie's wave function for a single free particle}

Now let us write down the complex-valued spacetime wave function $\psi(q;\tau)=I\exp(iS/\hbar)$ corresponding to the spacetime soliton that we have just developed. To do this, one has to calculate the quantum phase by integrating $\partial^aS=p^a$ to give us
\begin{equation}
S(q;\tau)=-\frac{E}{c}(ct)+p_xx+p_yy+p_zz+\xi(\tau),
\end{equation}
where $\xi(\tau)$ is a function only of $\tau$. Hence, at proper time $\tau$, one gets
\begin{eqnarray}
\psi(q;\tau)=I_{{\bf x}_0}({\bf x}-{\bf v}\tau)I_{t_0}(ct-E\tau/(mc))\hspace{23mm}\nonumber\\
\times\exp\Big(i(-Et+p_xx+v_yy+p_zz+\xi(\tau))/\hbar\Big). 
\label{de Broglie wave function 1}
\end{eqnarray}
Inserting this into the relativistic Schr\"odinger equation of Eq. (\ref{covariant Schroedinger equation}) one can show that $\xi(\tau)$ is related to $\langle P\rangle$ as $d\xi/d\tau=-\langle P\rangle$ which can be integrated to give $\xi(\tau)=-\langle P\rangle\tau=mc^2\tau$ up to some constant. Putting this back into Eq. (\ref{de Broglie wave function 1}) one finally obtains
\begin{eqnarray}
\psi(q;\tau)=I_{{\bf x}_0}({\bf x}-{\bf v}\tau)I_{t_0}(ct-E\tau/(mc))\hspace{23mm}\nonumber\\
\times\exp\Big(i(-Et+p_xx+p_y y+p_z z+mc^2\tau)/\hbar\Big). \nonumber\\ 
\label{de Broglie wave function 2}
\end{eqnarray}
One concludes that the phase of the soliton wave function uniquely gives the mass-energy-momentum of the  particle.

One can also see from Eq. (\ref{de Broglie wave function 2}) that the soliton wave function possesses internal vibration in spacetime whose wave number and angular frequency are given by the four momentum vector $p^a$ as 
\begin{equation}
{k_B}_i=p_i/\hbar, \hspace{2mm}i=x,y,z,\hspace{2mm}\mbox{and}\hspace{2mm}\omega_B=E/\hbar. 
\label{de Broglie guiding principle}
\end{equation}
The above relations are nothing but the de Broglie's conjecture in his attempt to explain the dual nature of matter as both particle and wave by relating the particle properties (energy-momentum) of the matter to the wave properties (angular frequency-wave number) of the matter. Yet, in contrast to our approach which is based on {\it linear wave equation}, de Broglie envisioned such soliton solution to be derived from a {\it nonlinear wave equation}. 

Now, let us assume that the internal vibration resonates inside the spatio-temporal support. Namely,  the wave number of the internal vibration are equal to the integer multiple of the wave number of the spatial part of quantum amplitude; and moreover, the angular frequency of the internal vibration is equal to the integer multiple of the angular frequency of the temporal part of the quantum amplitude:
\begin{eqnarray}
{k_B}_i=n_{i}k_{i_0},\hspace{2mm}i=x,y,z,\nonumber\\
\omega_B=n_t\omega_0. \hspace{20mm}
\label{resonance}
\end{eqnarray}
where $n_x,n_y,n_z,n_t$ are integer. Multiplying both sides of the above equations with Planck constant $\hbar$, one gets
\begin{eqnarray}
p_i=\hbar{k_B}_i=n_{i}\hbar k_{i_0},\hspace{2mm}i=x,y,z,\nonumber\\ 
E=\hbar \omega_B=n_t\hbar \omega_0. \hspace{20mm}
\label{Planck-Einstein quantization}
\end{eqnarray} 
The above obtained relations tells us that the energy-momentum are quantized into discrete values. 

Further, let us proceed to discuss a special case when $n_x=n_y=n_z=n_t=1$ to have
\begin{equation}
E=\hbar\omega_0, \hspace{2mm} {\bf p}=\hbar {\bf k}_{{\bf x}_0}.
\label{first resonance}
\end{equation}
Using the above relation between the energy-momentum of the soliton and the angular frequency-wave number of the spacetime quantum amplitude, then Eq. (\ref{energy-momentum-mass relation}) translates into:
\begin{equation}
\frac{E^2}{c^2}-{\bf p}^2=m^2c^2.
\label{energy-momentum relation}
\end{equation}
This is the classical energy-momentum relation of special relativity. Since in general one can choose ${\bf k}_{{\bf x}_0}$ and $\omega_0$ independently of ${\bf p}$ and $E$, then the above result opens the possibility of the violation of classical energy-momentum relation provided that Eq. (\ref{first resonance}) is not satisfied.   

It is then interesting to ask when the resonance of the internal spatio-temporal vibration inside the spacetime support occurs. To answer this, it is intuitive to learn from the classical wave phenomena. In this case, the resonance phenomena occurs if there is interaction and transfer of energy. One therefore might guess that the quantization of energy-momentum of Eqs. (\ref{Planck-Einstein quantization}) occurs if there is interaction and transfer of energy-momentum. In other words, we expect that the energy-momentum can only be transferred from one matter to the other through interaction in discrete quantized value. For the case of light, this statement is but the Planck-Einstein conjecture which eventually gave birth to quantum theory. To discuss this issue in precise manner, one has to develop a theory describing interaction among particles. 

\subsection{Spacetime uncertainties}

One of the important property of the spacetime soliton wave function we developed in the previous subsections is that it is  broadened with finite width both in space and time axes. Namely, any particle should be considered as an extended object both in space and time. One is thus suggested to abandon the view of seeing a particle as a {\it dimensionless point} except for approximation to certain situation. 

It is well-known that the assumption of point particle has led to many formal and physical difficulties in physics. It is for example argued as the origin of the divergence of calculations in the current version of quantum field theory.  Another example is the famous paradox of self-interference in double slits experiment \cite{Bell unspeakable}. Assuming a point particle will force one to say an ambiguous sentence that {\it the particle is passing through both slits to interfere with itself at the screen}. Our soliton model of particle developed in this paper might then be considered as prospective candidate to address these important  issues. It is thus imperative to understand how the width of the soliton wave function is related to other directly observable physical quantities. 

First, let us discuss the spatial width of the soliton wave function defined by $\Delta_i=2i_0$, $i=x,y,z$. Recalling the relation $k_{i_0}i_0=\pi/2$, $i=x,y,z$ of Eq. (\ref{wave number vs support}), which is coming from the boundary condition at the blowing-up point, one has 
\begin{equation}
\Delta_i=\frac{\pi}{k_{i_0}},\hspace{2mm}i=x,y,z.
\label{spatial width of soliton}
\end{equation}
Hence the width in the $i-$axis is proportional to the inverse of the $i-$part of the wave number of the spatial quantum amplitude. To see its relation with directly observable quantities, it is convenient to define the following quantity:
\begin{eqnarray}
\Delta_{\bf x}\equiv\frac{\Delta_x\Delta_y\Delta_z}{\sqrt{(\Delta_y)^2(\Delta_z)^2+(\Delta_x)^2(\Delta_z)^2+(\Delta_x)^2(\Delta_y)^2}}\nonumber\\
=\frac{\pi}{\sqrt{k_{{\bf x}_0}^2}}=\frac{\pi\hbar}{\sqrt{\frac{\hbar^2\omega_0^2}{c^2}-m^2c^2}}. \hspace{20mm}
\end{eqnarray}
It depends on the angular frequency of the quantum amplitude and also on mass. In the case when the internal vibration resonates inside the spatio-temporal support in its first mode, the denominator is equal to the absolute value of the momentum of the particle to give:
\begin{equation}
\Delta_{\bf x}=\frac{\pi\hbar}{\sqrt{\frac{E^2}{c^2}-m^2c^2}}=\frac{\pi\hbar}{\|{\bf p}\|}. 
\label{de Broglie wave length}
\end{equation}

Next, let us discuss the temporal width of the soliton wave function defined similarly by: $\Delta_t\equiv 2t_0$. Again, using the relation $\omega_0t_0=\pi/2$ of Eq. (\ref{angular frequency vs support}), one gets
\begin{equation}
\Delta_t=\frac{\pi}{\omega_0}=\frac{\pi\hbar}{c\sqrt{\hbar^2k_0^2+m^2c^2}}. 
\label{time uncertainty}
\end{equation}
Similarly, notice that if the internal vibration beats inside the spatio-temporal support in its first mode, the denominator is just equal to the energy $E$ to give us 
\begin{equation}
\Delta_t=\frac{\pi\hbar}{E}
\end{equation} 

\section{Superposition of masses}

Now let us proceed to discuss the implication of the linearity of the Schr\"odinger equation of Eq. (\ref{covariant Schroedinger equation}). Recall that the soliton wave function given in Eq. (\ref{de Broglie wave function 2}) can be interpreted as the wave function of a single free particle with a given mass $m$, energy $E$ and momentum ${\bf p}=\{p_x,p_y,p_z\}$. Since the Schr\"odinger equation of Eq. (\ref{covariant Schroedinger equation}) is linear  with respect to the wave function, then any superposition of the solutions of the type given in Eq. (\ref{de Broglie wave function 2}) will also satisfy the Schr\"odinger equation. 

For example, let us assume that $\psi_i(q;\tau;\{m_i,E_i,{\bf p}_i\})$, $i=1,2$ takes the soliton form given in Eq (\ref{de Broglie wave function 2}) with the mass $m_i$, energy-momentum $\{E_i,{\bf p}_i\}$, $i=1,2$. Then, the following wave function which is obtained by superposing the two solitons wave functions:
\begin{eqnarray}
\psi(q;\tau)=A_1\psi_1(q;\tau;\{m_1,E_1,{\bf p}_1\})\hspace{10mm}\nonumber\\
+A_2\psi_2(q;\tau;\{m_2,E_2,{\bf p}_2\}),
\label{superposition of mass-wave function}
\end{eqnarray}
also satisfies the Schr\"odinger equation of Eq. (\ref{covariant Schroedinger equation}).  Here $A_i$, $i=1,2$ are two complex-valued constants which satisfy: $|A_1|^2+|A_2|^2=1$. One can thus interpret each soliton term on the right hand side as describing a particle with various values of masses, energies and momenta. 

Notice that at period of time when the spatio-temporal support of both solitons wave functions are overlapping, one will observe interference in space and time. Hence it is impossible to distinguish one soliton from the other. One therefore should consider both as a single particle. Now let us assume that ${\bf p}_1=-{\bf p}_2$, namely each soliton is moving with direction opposite to the movement of the other.  Then at sufficiently large time $\tau$, the two solitons are no more overlapping in space so that one can already distinguish one from the other. Both might however still overlap in time domain. If further one chooses $E_1/m_1\neq E_2/m_2$, then at sufficiently large proper time, both soliton will also separate in time axis. 

Notice also that both solitons move independently from the other. One can thereby calculate the total mass, $m_{12}$, to obtain
\begin{equation}
m_{12}=\sqrt{\frac{-2{\bar U}}{c^2}}=m_1+m_2.
\label{total mass-energy-momentum}
\end{equation}
This can be shown easily by choosing proper time $\tau$ so that both solitons are not overlapping in spacetime so that one has ${\bar U}={\bar U}_1+{\bar U}_2$, where ${\bar U}_i$ is the average quantum potential of soliton $i=1,2$. Hence, the manifestly covariant Schr\"odinger equation of Eq. (\ref{covariant Schroedinger equation}) can be seen as field theory in which a particle of mass $m_{12}$ can break into two particles of masses $m_1$ and $m_2$ so that $m_1+m_2=m_{12}$. Each particle moves independently with momentum ${\bf p}_i$ and possessing energy $E_i$, where $i=1,2$. Moreover, since energy and momentum must be conserved then one can define the total energy-momentum as $E_{12}=E_1+E_2$ and ${\bf p}_{12}={\bf p}_1+{\bf p}_2$.  Conversely, since the Schr\"odinger equation of Eq. (\ref{covariant Schroedinger equation}) is time reversal, then by reversing the evolution of affine parameter, the initially separated two solitons each describing a particle of mass $m_i$, $i=1,2$, can merge into a single particle with mass $m_{12}=m_1+m_2$. Needless to say, extension into the superposition of more than two solitons are straight forward. 

\section{Conclusion and Discussion}

We have thus shown that  first, the manifestly covariant mass-less Schr\"odinger equation of Eq. (\ref{covariant Schroedinger equation})  possesses a class of non-dispersive soliton solutions with finite-size spatio-temporal support.  The quantum amplitude inside the support satisfies Klein-Gordon equation with finite emergent mass. We finally came to give an interpretation to the soliton solution as describing a single particle with finite mass, energy and momentum. Moreover, we showed that the soliton solution possesses internal spatio-temporal vibration with angular frequency and wave number which are determined by the energy-momentum of the particle as exactly conjectured by de Broglie while proposing the solution for the problem of duality of matter as both particle and wave. However, in contrast to his envision of {\it nonlinear wave equation} assuming soliton solution, we showed that our soliton wave function is a solution of a {\it linear wave equation}. We argued that the fact that the soliton wave function has finite extension in spacetime might give the key to the problematic divergence problem encountered in the current version of quantum field theory and also to the paradox of particle self-interference in double slits experiment. 

We showed that if the internal vibration resonates inside the spatio-temporal support, then one can show that the energy and momentum are discretized into packets as assumed by Planck and Einstein. We further suggested that this situation appears if there is interaction among matters which induces transfer of energy. We also showed that the classical energy-momentum relation is recovered when the internal vibration resonates inside the spatio-temporal support in its first excited mode. This opens the possibility of violating the energy-momentum relation if one moves away from the first resonance mode.  

Next, the linearity of the Schr\"odinger equation allows for superposition of such solutions to further comprise a class of many solitons solutions. Each term of the superposition then describes a single particle with a given mass, energy and momentum, moving independently of the other. One can thus consider the wave function as a real physical field and regards the Schr\"odinger equation of Eq. (\ref{covariant Schroedinger equation}) as a field theory describing multi-particles systems. In this theory, things thus live in the ordinary three dimensional space rather than in configuration space. 

There are at least three other interesting issues suggest further exploration. First, we have identified soliton as particle by attributing to it mass and energy-momentum. It is then natural to ask: what about spin? Second, since the soliton wave function is broadened also in time axis, then it might be possible to see interference in the time domain \cite{Lindner interference in time}. Finally third, our theory allows us to construct a superposition of two solitons which initially interferes each other but  then is separated from each other as time goes. It is then interesting to employ this idea to re-think the EPR thought experiment. 

\begin{acknowledgments}
This work is supported by FPR program at RIKEN. 
\end{acknowledgments}


\begin{thebibliography}{10}

\bibitem{Penrose book}Roger Penrose, {\it The large, the small and human mind} (Cambridge University Press, Cambridge, 2000). 

\bibitem{Isham book} C. J. Isham, {\it Lectures On Quantum Theory: Mathematical and Structural Foundation} (Imperial College Press, London, 1995). 

\bibitem{Bell unspeakable}J. S. Bell, {\it Speakable and Unspeakable in Quantum Mechanics} (Cambridge University Press, Cambridge, 2004).

\bibitem{Bohm-Hiley book}D. Bohm and B. J. Hiley,  {\it The Undivided Universe: An ontological interpretation of quantum theory} (Routledge, London, 1993). 

\bibitem{Born paper}M. Born Z. Phys. {\bf 37}, 863 (1926).

\bibitem{Dirac book}P. A. M. Dirac, {\it The Principle of Quantum Mechanics} (Clarendon Press, UK, 1981). 

\bibitem{von Neumann book}John von Neumann, {\it Mathematical Foundation of Quantum Mechanics} (Princeton University Press, Princeton, 1996). 

\bibitem{Bohm paper}D. Bohm, Phys. Rev. \textbf{85}, 166 (1952); \textbf{85}, 180 (1952); \textbf{89}, 458 (1953). 

\bibitem{Everett many worlds}H. Everett, Rev. Mod. Phys. {\bf 29}, 454 (1957). 

\bibitem{deWitt and Graham book}B. S. DeWitt and N. Graham, {\it The Many-Worlds Interpretation of Quantum Mechanics} (Princeton University Press, Princeton, 1973). 

\bibitem{GRW}G. C. Ghirardi, A. Rimini, and T. Weber,  Phys. Rev. D {\bf 34},  470 (1986). 

\bibitem{de Broglie book} L. de Broglie, {\it Une tentative d'interpretation causale et nonlineaire de la mecanique ondulatoire} (Gauthier-Villars, Paris, 1956). 

\bibitem{de Broglie late book}L. de Broglie, {\it Heisenberg's Uncertainties and the Probabilistic Interpretation of Wave Mechanics} (Kluwer Academic Publisher, London, 1990). 

\bibitem{Pauli book}W. Pauli, {\it General principles of quantum mechanics} (Springer, Berlin, 1980). 

\bibitem{Curfaro-Petroni paper} N. Curfaro-Petroni {\it et. al.}, Phys. Rev. D {\bf 32}, 1375 (1985).

\bibitem{Vigier paper 1} J. P. Vigier, Astron. Nachr. {\bf 303}, 1 (1982). 

\bibitem{Vigier paper 2} J. P. Vigier, Phys. Lett. A {\bf 135}, 99 (1989).

\bibitem{Mackinnon paper 1} L. Mackinnon, Found. Phys. {\bf 8}, 157 (1978).

\bibitem{Barut paper}A. O. Barut, Phys. Lett. A {\bf 143}, 349 (1990).  

\bibitem{Nambu}Y. Nambu, Prog. Theor. Phys. {\bf 5}, 82 (1950). 

\bibitem{Kyprianidis}A. Kyprianidis, Phys. Rep. {\bf 155}, 1 (1987).

\bibitem{Madelung paper}E. Madelung, Zeits. F. Phys. \textbf{40}, 332 (1926).

\bibitem{AgungPRA1}Agung Budiyono and Ken Umeno, Phys. Rev. A \textbf{79}, 042104 (2009).

\bibitem{blowing-up NDE}S. Alinhac, {\it Blowup for Nonlinear Hyperbolic Equations} (Birkh\"auser, Basel, 1995). 

\bibitem{Lindner interference in time}F. Lindner, M. G. Sch\"atzel, H. Walther, A. Baltuska, E. Goulielmakis, F. Krausz, D.B. Milosevic, D. Bauer, W. Becker and G.G. Paulus, Phys. Rev. Lett. {\bf 95}, 040401 (2005).

\end{thebibliography}
\end{document}